\documentclass{article}
\usepackage{PdMaxCon25template}
\usepackage{times}
\usepackage{ifpdf}
\usepackage{soul}
\usepackage{booktabs}
\usepackage{siunitx}
\usepackage[english]{babel}
\newcommand{\tildeobj}{\texttt{\raisebox{.3ex}{\texttildelow}}}

\usepackage{amsmath} 
\usepackage{amssymb} 
\usepackage{amsfonts}
\usepackage{bm}      

\def\papertitle{nlm: Real-Time Non-linear Modal Synthesis in Max}
\def\firstauthor{Rodrigo Diaz}
\def\secondauthor{Rodrigo Constanzo}
\def\thirdauthor{Mark Sandler}

\newif\ifpdf
\ifx\pdfoutput\relax
\else
   \ifcase\pdfoutput
      \pdffalse
   \else
      \pdftrue
  \fi
\fi

\ifpdf 
  \usepackage[pdftex,
    pdftitle={\papertitle},
    pdfauthor={\firstauthor, \secondauthor, \thirdauthor},
    bookmarksnumbered, 
    pdfstartview=XYZ 
   ]{hyperref}

  \usepackage[pdftex]{graphicx}
  \graphicspath{{./figures/}}
  \DeclareGraphicsExtensions{.pdf,.jpeg,.png}

  \usepackage[figure,table]{hypcap}

\else 
  \usepackage[dvips,
    bookmarksnumbered, 
    pdfstartview=XYZ 
  ]{hyperref}  

  \usepackage[dvips]{epsfig,graphicx}
  \graphicspath{{./figures/}}
  \DeclareGraphicsExtensions{.eps}

  \usepackage[figure,table]{hypcap}
\fi

\hypersetup{
    colorlinks,%
    citecolor=black,%
    filecolor=black,%
    linkcolor=black,%
    urlcolor=black
}

\usepackage{cleveref}

\title{\papertitle}

 \threeauthors
   {0.2in}
   {\firstauthor} {Queen Mary University of London \\ %
     {\tt \href{mailto:r.diazfernandez@qmul.ac.uk}{r.diazfernandez}}}
   {\secondauthor} {Royal Northern College of Music \\ %
     {\tt \href{mailto:rodrigo.constanzo@rncm.ac.uk}{rodrigo.constanzo}}}
   {\thirdauthor} {Queen Mary University of London \\ %
     {\tt \href{mailto:mark.sandler@qmul.ac.uk}{mark.sandler}}}

\begin{document}
\capstartfalse
\maketitle
\capstarttrue
\begin{abstract}
\sloppy
We present \texttt{nlm}, a set of Max externals that enable efficient real-time non-linear modal synthesis for strings, membranes, and plates. The externals, implemented in C++, offer interactive control of physical parameters, allow the loading of custom modal data, and provide multichannel output. By integrating interactive physical-modelling capabilities into a familiar environment, \texttt{nlm} lowers the barrier for composers, performers, and sound designers to explore the expressive potential of non-linear modal synthesis. The externals are available as open-source software at \url{https://github.com/rodrigodzf/nlm}.
\end{abstract}

\section{Introduction}
\label{sec:introduction}

Physical modelling synthesis is a technique for creating sound by simulating the physical properties of real-world objects. It is a powerful tool for creating realistic and complex sounds, and is used in a wide range of applications, from music composition to sound design. 

Several numerical approaches have been used to implement physical modelling synthesis, including Finite Difference Time Domain (FDTD) methods~\cite{bilbao_numerical_2009}, Digital Waveguide Meshes (DWM), and a broad family of techniques often grouped under the term \emph{modal methods}. These include classical spectral and pseudospectral approaches~\cite{bilbao_modaltype_2004}, the Functional Transformation Method (FTM)~\cite{trautmann_functional_2003} and to an extent the Finite Element Method (FEM). All of these techniques share the property of decomposing the system into modes that can evolve independently in the linear case and interact in the presence of non-linearities. Due to their compact representation, physical interpretability, and efficient time integration, modal methods are especially attractive in environments like Max and Pure Data (Pd), where real-time interaction is a core requirement.

In this work, we present \texttt{nlm}, a unified and optimised set of Max externals implementing non-linear modal synthesis models previously described in the literature~\cite{trautmann_functional_2003, avanzini_efficient_2012, ducceschi_simulations_2015}. While these non-linear modal models have been extensively studied and implemented in offline simulations, their deployment within Max has remained limited. Our contribution is the creation of a unified and computationally efficient real-time implementation, using the Eigen~\cite{guennebaud_eigen_2010} C++ library for matrix operations. The externals are designed to allow interactive exploration of the parameters of the system, with support for user-defined eigenmodes, and creative exploration of excitation profiles.

\section{Background}
\label{sec:background}

\subsection{Modal Synthesis in Max and Pd}
\label{subsec:modal_synthesis_in_max_pd}

Modal methods have long been central to the development of real-time physical modelling tools within Max and Pd. Their decomposition into resonant modes naturally aligns with modular software architectures, making them particularly well-suited for real-time interactive synthesis. In this section, we provide a short overview of influential modal synthesis tools developed specifically for these environments, focusing on those with notable impact on the field.

Early contributions to Max include the \emph{NVM} externals~\cite{dudas_nvma_1998}, a modular physical modelling suite demonstrating real-time performance with hundreds of modes. Later, IRCAM introduced the \texttt{mlys} and \texttt{modalys\tildeobj} objects, integrating the \emph{Modalys} synthesis framework into Max and enabling users to define complex resonator networks interactively~\cite{causse_modalys_2011}. Concurrently, Trueman and DuBois developed the \emph{PeRColate} collection~\cite{trueman_percolate_2015}, wrapping waveguide and modal components from the Synthesis Toolkit~\cite{scavone_rtmidi_2005} into Max objects, facilitating synthesis of percussive and other instrument types.

More recently, the \emph{Sound Design Toolkit} (SDT)~\cite{baldan_sound_2017} provided a cross-platform C core compiled into externals for both Max and Pd. The \texttt{sdt} objects support modal resonators which can be excited by impact, friction, and other physical interactions, allowing efficient simulation of complex non-linear collisions. Similarly, the \emph{Synth-A-Modeler} compiler~\cite{berdahl_introduction_2012} translates FAUST diagrams into optimised externals for Max, offering a high-level prototyping approach for modal synthesis. The implemented resonators in these cases, however, have been limited to linear modal synthesis.


While non-linear plate models, such as VKGong\footnote{\url{https://vkgong.ensta-paris.fr}} and VKPlate\footnote{\label{vkplate_repo}\url{https://github.com/Nemus-Project/VKPlate}}, have been implemented in C++ and MATLAB, real-time implementations of these non-linear models have not yet been fully integrated into Max or Pd.

\subsection{Non-linear Modal Models}
\label{subsec:non_linear_modal_models}
Assuming isotropic, homogeneous material properties, the dynamics of non-linear strings, membranes, and plates with bending stiffness can be described by the general scalar partial differential equation:
\begin{equation}
\label{eq:wave_equation}
\rho \ddot{w} + \left(d_1 + d_3 \Delta\right)\dot{w} + (D \Delta \Delta - T_0 \Delta) w = f_{\text{ext}} - f_{\text{nl}},
\end{equation}
where $w(\mathbf{x}, t)$ is transverse displacement, $\rho$ is the material density (i.e., $\rho = \rho_{\text{m}} A$ in the string case and $\rho = \rho_{\text{m}}h$ in the other cases), $d_1, d_3$ are the damping coefficients, $D$ is the bending stiffness, and $T_0$ is the initial tension. The $\Delta$ and $\Delta \Delta$ operators are the Laplacian and biharmonic operators respectively, and the terms $f_{\text{ext}}$ and $f_{\text{nl}}$ represent external and non-linear forces. The parameters with their units are summarised in~\Cref{tab:params}.

\Cref{eq:wave_equation} and its modal formulation have previously been presented and discussed in the literature for strings, membranes~\cite{trautmann_functional_2003, avanzini_efficient_2012} and plates~\cite{ducceschi_simulations_2015}. Here we provide a streamlined, unified presentation to ensure the paper remains self-contained.

Non-linearities in these models arise from deformation induced tension, in the Kirchhoff-Carrier model for strings and the Berger approximation for membranes. For plates, described by the von Kármán model, non-linearities emerge through coupling between transverse and in-plane modes.

The displacement can be decomposed as a modal expansion:
\begin{equation}
\label{eq:modal_expansion}
w(\mathbf{x}, t) = \sum_{\mu=1}^{M} \frac{\Phi_{\mu}(\mathbf{x}) q_{\mu}(t)}{\left\|\Phi_{\mu}(\mathbf{x})\right\|^2},
\end{equation}
where each $\Phi_{\mu}, \lambda_{\mu}$ eigenpair satisfies the eigenvalue problem $\Delta \Phi_{\mu} = -\lambda_{\mu} \Phi_{\mu}$. Analytical solutions are limited to simple geometries and boundary conditions, and more complex cases require numerical approaches. In the von Kármán case, an analogous expansion of the Airy stress function yields in-plane eigenpairs \((\Psi_\nu,\zeta_\nu)\).

Truncating the modal expansion to $M$ modes, the system is described by the modal coordinates $\mathbf{q} = [q_{\mu}]_{\mu=1}^{M} \in \mathbb{R}^{M}$. Substituting into~\eqref{eq:wave_equation} yields:
\begin{equation}
\label{eq:modal_ode}
\ddot{q}_{\mu} + 2\gamma_{\mu}\dot{q}_{\mu} + \omega_{\mu}^2 q_{\mu} = \bar{f}_{\text{ext},\mu} - \bar{f}_{\text{nl},\mu},
\end{equation}
with
\begin{align}
\omega_{\mu}^2 &= \frac{D\lambda_{\mu}^2 + T_0\lambda_{\mu}}{\rho}, \quad \gamma_{\mu} = \frac{d_1 + d_3\lambda_{\mu}}{2\rho}.
\end{align}
Applying the impulse-invariant discretisation to~\eqref{eq:modal_ode} yields the explicit update scheme:
\begin{equation}
\mathbf{q}^{n+1} = \mathbf{a}_1 \odot \mathbf{q}^n + \mathbf{a}_2 \odot \mathbf{q}^{n-1} + \mathbf{b}_1 (\mathbf{\bar{f}}_{ext}^n - \mathbf{\bar{f}}_{nl}^n),
\end{equation}
where $\odot$ denotes element-wise multiplication of vectors, and the coefficient vectors are defined as:
\begin{align}
    \mathbf{a}_1 &= -2 e^{-\boldsymbol{\gamma}T} \odot \cos(\tilde{\boldsymbol{\omega}}T), \\
    \mathbf{a}_2 &= e^{-2\boldsymbol{\gamma}T}, \\
    \mathbf{b}_1 &= \frac{1}{\rho\tilde{\boldsymbol{\omega}}} \odot \sin(\tilde{\boldsymbol{\omega}}T) \odot e^{-\boldsymbol{\gamma}T},
\end{align}
with $\tilde{\boldsymbol{\omega}} = \sqrt{\boldsymbol{\omega}^2 - \boldsymbol{\gamma}^2}$ and sampling period $T$. A similar second-order update scheme can be derived using centered finite differences (St\"ormer-Verlet)~\cite{ducceschi_simulations_2015}. In modal coordinates, the non-linear terms for string, membrane, and plate models become:
\begin{align}
\bar{f}_{\mu,\text{string-nl}} &= \lambda_\mu q_\mu \frac{EA}{2L \rho} \sum_\mu \frac{\lambda_\mu q_\mu^2}{\left\|\Phi_\mu\right\|^2}, \label{eq:string-nl} \\
\bar{f}_{\mu,\text{membrane-nl}} &= \lambda_\mu q_\mu \frac{Eh}{2L_x L_y (1-\nu^2)\rho} \sum_\mu \frac{\lambda_\mu q_\mu^2}{\left\|\Phi_\mu\right\|^2} \label{eq:membrane-nl}, \\
\bar{f}_{\mu,\text{plate-nl}} &= \frac{E}{2 \rho_{\text{m}}} \sum_{p, q, r}^n \frac{H_{q, r}^n C_{p, n}^s}{\zeta_n^4} q_p q_q q_r, \label{eq:plate-nl}
\end{align}
where $H_{q, r}^n$ and $C_{p, n}^s$ are third-order tensors containing coupling coefficients, where $p,q,r$ index the transverse and $n,s$ index the in-plane modes, respectively. More details on the derivation of~\Cref{eq:string-nl,eq:membrane-nl,eq:plate-nl} can be found in~\cite{ducceschi_simulations_2015,avanzini_efficient_2012}.

\section{Max Implementation}
\label{sec:max_implementation}

The package consists of a C++ core library and four externals: \texttt{nlm.string\tildeobj}, \texttt{nlm.plate\tildeobj}, \texttt{mcs.nlm.string\tildeobj}, \texttt{mcs.nlm.plate\tildeobj}, where the last two are multi-channel versions. Given the similarity between the membrane and plate models, both are encapsulated within a single external \texttt{nlm.plate\tildeobj}. The type of model can be selected via an attribute, specified as \texttt{berger} for the membrane or \texttt{vk} for the von Kármán plate model. The string and membrane implementations follow previous work~\cite{trautmann_functional_2003,avanzini_efficient_2012}, while the calculation of coupling coefficients for the plate follows~\cite{ducceschi_simulations_2015} and the MATLAB implementation provided by the VKGong project.

Both externals expose the physical parameters of the governing equation~\eqref{eq:wave_equation} as interactive attributes, summarised in~\Cref{tab:params}. The externals currently support only rectangular geometries with simply supported boundary conditions (i.e., zero displacement and bending at the edges). However, users can optionally load custom mode shapes, eigenvalues, and modal coupling coefficients that have been computed externally using other numerical methods, enabling simulations of more complex geometries or alternative boundary conditions. Since these coefficients depend on geometry, loading custom eigenpairs disables runtime geometric attributes. To support practical use, help files and example presets with a selection of material/geometry configurations are included.

\begin{table}[h]
    \centering
    \resizebox{\columnwidth}{!}{%
    \begin{tabular}{@{}llcc@{}}
    \toprule
    \textbf{Parameter} & \textbf{Attribute} & \textbf{String} & \textbf{Plate} \\
    \midrule
    $E$ & \texttt{youngs\_modulus} & \multicolumn{2}{c}{\si{\giga\pascal}} \\
    $\rho_{\text{m}}$ & \texttt{density} & \multicolumn{2}{c}{\si{\kilo\gram\per\metre\tothe{3}}} \\
    $L$, $L_x$, $L_y$ & \texttt{lx}, \texttt{ly} & \multicolumn{2}{c}{\si{\metre}} \\
    $h$ & \texttt{thickness} & — & \si{\metre} \\
    $A$ & \texttt{cross\_sectional\_area} & \si{\milli\metre\squared} & — \\
    $I$ & \texttt{moment\_of\_inertia} & \si{\milli\metre\tothe{4}} & $\frac{h^3}{12}$ \\
    $\nu$ & \texttt{poisson\_ratio} & — & — \\
    $T_0$ & \texttt{tension} & \si{\newton} & \si{\newton\per\metre} \\
    $d_1$ & \texttt{f\_independent\_loss} & \si{\kilo\gram\per\metre\per\second} & \si{\kilo\gram\per\metre\squared\per\second} \\
    $d_3$ & \texttt{f\_dependent\_loss} & \si{\kilo\gram\metre\squared\per\second} & \si{\kilo\gram\per\second} \\
    $D$ & — & $EI$ & $\frac{Eh^3}{12(1-\nu^2)}$ \\
    \bottomrule
    \end{tabular}%
    }
    \caption{Physical parameters for string and membrane/plate models and their attributes in Max. We use moment of inertia following the definitions in~\cite{trautmann_functional_2003}. For the string $I = \kappa^2 A$ where $\kappa$ is the radius of gyration. Note that the \texttt{density} attribute for the externals is the mass density, not the material density used in~\eqref{eq:wave_equation}. The bending stiffness $D$ is computed internally and it is not exposed as an attribute directly.}
    \label{tab:params}
\end{table}
    
\subsection{Interaction}
\label{subsec:interaction}

\begin{figure}[h]
    \centering
    \includegraphics[width=\columnwidth]{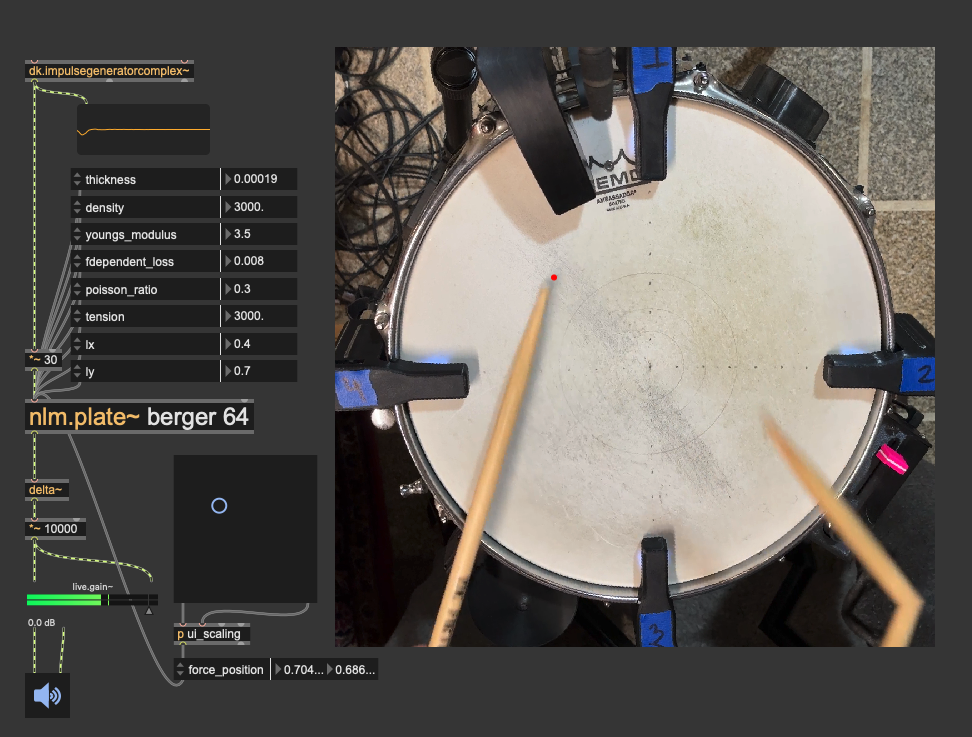}
    \caption{Screenshot of the \texttt{nlm.plate\tildeobj} external using the \texttt{dk.impulsegeneratorcomplex} and live input.}
    \label{fig:nlm_plate}
\end{figure}

In the current implementation, we assume that the external force applied to the system is independent of the resonator state. That is, it does not depend on the displacement or velocity of the string or plate at the point of contact. This simplifies the interaction model by allowing externally defined force profiles (e.g., envelopes, filtered noise, or live inputs) to be applied directly without requiring a dynamic feedback loop between the input and the state of the resonator. When a force $f_{\text{exc}}(t)$ is applied at a specific point $\mathbf{x}_0$ on the structure, its modal contribution is computed as:
\begin{equation}
\label{eq:modal_force}
\bar{f}_{\mathrm{ext}, \mu}(t) = \phi_\mu(\mathbf{x}_0) \cdot f_{\text{exc}}(t)
\end{equation}
The excitation signal position $\mathbf{x}_0$ can be defined flexibly. For example, by using sensor data from a drum head or a string instrument. In a more musical context, we have explored a hybrid excitation method in which the input force signal is composed of two components: a filtered white noise signal and a filtered signal derived from a live recording. This approach has been implemented in the \texttt{dk.impulsegeneratorcomplex} object, available as part of the SP-Tools package\footnote{\url{https://github.com/rconstanzo/SP-tools}}. Although this excitation strategy does not strictly adhere to physical modelling of the external force, it offers enhanced creative possibilities for sound design. This is particularly valuable given that the primary audience for this object includes musicians and sound designers seeking intuitive and creative control. A screenshot of the \texttt{nlm.plate\tildeobj} external using the positional excitation tracking and the hybrid excitation method is shown in~\Cref{fig:nlm_plate}.

\section{Conclusions}

We have presented \texttt{nlm}, a unified set of Max externals providing an efficient real-time implementation of non-linear modal synthesis models for strings, membranes, and plates. Developed in C++ and using Eigen optimisations, our externals offer computational performance suitable for interactive audio applications. The externals also allow interactive parameter configuration, the optional loading of custom eigenpairs for arbitrary geometries, and multichannel support for multiple readout positions. While only modal coupling matrices for rectangular geometries with simply supported boundary conditions are computed internally, users may optionally load externally pre-computed coupling matrices (calculated using external software) for more complex configurations. Overall, \texttt{nlm} lowers the barrier for musicians, composers, and sound designers to explore non-linear physical models within a familiar real-time audio environment.

Currently, one limitation of the implementation concerns numerical stability. While the linear modal filterbank is unconditionally BIBO-stable thanks to the impulse-invariant transform, this stability does not extend to the non-linear coupled system. In practice, we enforce the sampling-rate condition~\cite{ducceschi_simulations_2015}: $f_s > \omega_{\text{max}} / 2$, and the implementation remains numerically stable under typical performance conditions. However, sufficiently strong excitations can inject enough energy to result in numerical instability. To mitigate this, we plan to implement an energy-clamping strategy similar to that proposed for the slapped bass model~\cite{trautmann_functional_2003}, which scales the non-linear feedback force based on the instantaneous discrete energy.

Another practical limitation is computational load, which can rise sharply with the modal count. For the \texttt{vk} model, the per-sample update time-complexity is $O(N M^{2})$~\cite{ducceschi_simulations_2015}, where $N$ denotes the number of in-plane modes. Large values can overburden the CPU and produce audio clicks, especially when parameters are adjusted in real time. On current hardware, however, the externals run smoothly with roughly 100 plate modes (\texttt{vk}) and with several hundred modes for the simpler string and membrane models.

Several additional extensions are planned for future work. First, we aim to address the current instability in extreme input cases through more robust integration schemes. Second, we intend to expand the expressiveness of the system by incorporating non-linear contact models (similar to the ones available in the \texttt{SDT} package~\cite{baldan_sound_2017}). Third, although the current system already allows users to import modal data and coupling matrices for arbitrary geometries, we plan to provide native support for such cases to streamline the workflow and avoid external preprocessing. Finally, we also plan to release a Pd version of the externals.

Overall, our work offers a flexible and extensible set of tools for exploring the expressive potential of non-linear modal approaches.

\bibliography{references}

\end{document}